\documentclass[aps,prb,showpacs,twocolumn]{revtex4}
\usepackage{mathrsfs}
\usepackage{amssymb}
\usepackage{amsmath}
\usepackage{bm}
\usepackage{graphics}
\usepackage{natbib}

\newcommand{\sss}[1]{{\scriptscriptstyle #1}}
\begin{document}
\title{Electronic transport through a double-quantum-dot Aharonov-Bohm interference device with impurities}
\author{Wei-Jiang Gong$^{1,2}$}\email[Email
address: ]{weijianggong@gmail.com}
\author{Xue-Feng Xie$^1$}
\author{Yu Han$^3$}
\author{Guo-Zhu Wei$^{1,2}$}
\affiliation{1. College of Sciences, Northeastern University,
Shenyang 110004, China\\
2. International Center for Material Physics, Acadmia Sinica,
Shenyang 110015, China\\
3. Physics Department, Liaoning University, Shenyang 110015, China }
\date{\today}

\begin{abstract}
The impurity-related electron transport through a double quantum dot
(QD) Aharonov-Bohm (AB) interferometer is theoretically studied, by
considering impurities coupled to the QDs in the interferometer
arms. When investigating the linear conductance spectra \emph{vs}
the impurity levels, we show that the impurities influence the
electron transport in a nontrivial way, since their suppressing or
enhancing the electron tunneling. A presented single-level impurity
leads to the appearance of Fano lineshapes in the conductance
spectra in the absence of magnetic flux, with the positions of Fano
antiresonances determined by both the impurity-QD couplings and the
QD levels separated from the Fermi level, whereas when a magnetic
flux is introduced with the the phase factor $\phi=\pi$ the
impurity-driven Breit-Wigner lineshapes appear in the conductance
curves. Besides, the nonlocal impurities alter the period of
conductance change \emph{vs} the magnetic flux. The multi-level
impurities indeed complicate the electron transport, but for the
cases of two identical local impurities coupled to the respective
QDs with uniform couplings or a nonlocal impurity coupled to both
QDs uniformly, the antiresonances are only relevant to the impurity
levels. When many-body effect is managed within the second-order
approximation, we also find the important role of the Coulomb
interaction in modifying the electron transport.

\end{abstract}
\pacs{73.21.La, 73.63.-b, 73.23.Hk} \maketitle

\bigskip

\section{Introduction}
Current experimental developments and advances at the nanometer
scale have allowed to realize and manipulate QDs in a controlled
way\cite{Petta}, which motivated researchers to devote themselves to
the electron properties in QD systems with the observation of the
atom-like characteristics of a QD, such as the discrete electron
level and strong electron correlation\cite{Kool}, conductance
oscillation, and the Kondo effect\cite{Mahalu,Cronenwett,Heary}. And
these features permitted to employ QDs to build various
nano-devices, including rectifiers, amplifiers, lasers,
etc\cite{Ono,Vidan,Borri}.

\par

Due to their similitude with natural atoms, QDs are usually viewed
as artificial atoms, and then two or more coupled QDs are regarded
as artificial molecules. Multiple QD systems are of particular
interest and importance, since they possess higher freedom and admit
different kinds of connections to leads in implementing some
functions of quantum devices\cite{Ono,Vidan,Borri,Ustinov,Xie1}.
With respect to various multiple-QD structures, the configurations
of double QDs, i.e., the simplest coupled QDs, have received more
attention than the others both experimentally and
theoretically\cite{Fujisawa}. Initially, most of studies considered
serially-coupled double QDs\cite{Vaart}, but posteriorly the
T-shaped and parallel configurations were
examined\cite{Decker,Claro,Kang,Iye1,Iye2,T-shaped,PQD}, in which
the quantum interference are comparatively complicated due to the
existence of different transmission paths, and the peculiar electron
transport properties in them were thereby found, respectively. It is
known that for the structures of T-shaped QDs, when either
eigenlevel of the side-coupled QDs reaches the Fermi level of the
system the electron transmission is sharply suppressed which is the
so-called antiresonance phenomenon\cite{T-shaped}. As for the
parallel double-QD configurations, they, also described as the
double-QD molecule embedded in an AB ring, have also been
extensively investigated, where the presence of a magnetic flux adds
a new tool in the control of electron transport
behaviors\cite{PQD,Konig}. Accordingly, the AB oscillations have
been observed theoretically, accompanied by the support of lots of
experimental works. Then, the phenomenon of the effective
flux-dependent level attraction was also found.\cite{Konig} On the
other hand, in the cases of the appropriate QD-lead couplings, the
well-known Fano effect was examined, which motivated many studies
about this issue, including the tunable Fano lineshape by the
magnetic or electrostatic fields applied on the
QDs,\cite{refKang,refOrellana} the Kondo resonance associated Fano
effect,\cite{refKonig1,refBulka2} Coulomb-modification on the Fano
effect,\cite{refZhu2} the relation between the dephasing time and
the Fano parameter $q$\cite{refClerk}, and the spin-dependent Fano
effect associated with the Rashba spin-orbital (SO)
coupling.\cite{Sun,Chi}
\par
Despite these mentioned works, so far it is still a formidable
challenge to fabricate two clean QDs in the experiments due to
irregularities and defects in the QD system. Some localized states
often appear in QD systems, which are hybridized with the QD levels.
But they are not coupled directly to the leads, which, thereby, are
called the impurity states. In some previous literature, the effect
of impurities on the electron coherence transmission through the
low-dimensional systems has been
reported.\cite{BIC,Wu,PRL,EPL,NPJ,Jana,Riera,Faini,Qwire,Garmon}
Some researchers demonstrated the trapping of a conduction electron
between two identical impurities in a quantum wire (or a QD array),
as a result, the quantum interference was modified to a great extent
and there appear notable phenomena of bound states in
continuum.\cite{BIC,Qwire} Other groups showed that the
impurity-driven Fano resonances in quantum wires are sensitively
related to the parameters of impurity, such as its strength,
position, and lateral extent, which produces substantially different
effects on the Fano resonance. Wu $et$ $al.$ studied the
phonon-assisted electron transmission through a single QD with
impurity scattering effects.\cite{Wu} The results showed that
impurity states have weak influence on the suppression of the shot
noise. More recently, in a one-dimensional quantum wire
side-hybridized with a single-level impurity, the low-temperature
properties of interacting electrons were studied.\cite{PRL} It
reported that the hybridization induces a backscattering of
electrons in the wire which strongly affects its low-energy
properties. Thereby, one can predict that the impurity states play
an important role in the quantum transport process.
\par
With respect to such a topic, it is desirable to investigate the
influence of impurities on electron transport coupled-QD structures.
In the present paper, we study the quantum interference in electron
transport through a double QD AB interferometer modified by
impurities side-coupled to the QDs in its arms. As a result, it is
found the notable contribution of impurities to the electron
transmission through this structure. We show that the presence of a
single-level impurity induces the appearance of Fano lineshapes in
the conductance spectra in the zero-magnetic-field case, whereas it
drives the conductance spectra to exhibit Breit-Wigner lineshapes by
the presence of magnetic flux, in which the properties of the
corresponding lineshapes are strongly determined by the impurity-QD
couplings. Besides, the nonlocal impurities alter the period of
conductance change \emph{vs} the magnetic flux. The multi-level
impurities further transform the electron transport with the
complication of conductance curves, but for the cases of each QD
couples to the impurities uniformly, the antiresonances are only
relevant to the impurity levels. Also, we see the interesting effect
of the Coulomb interaction on the adjustment of the electron
traveling.

\section{The theoretical model\label{theory}}

\begin{figure}
\centering \scalebox{0.35}{\includegraphics{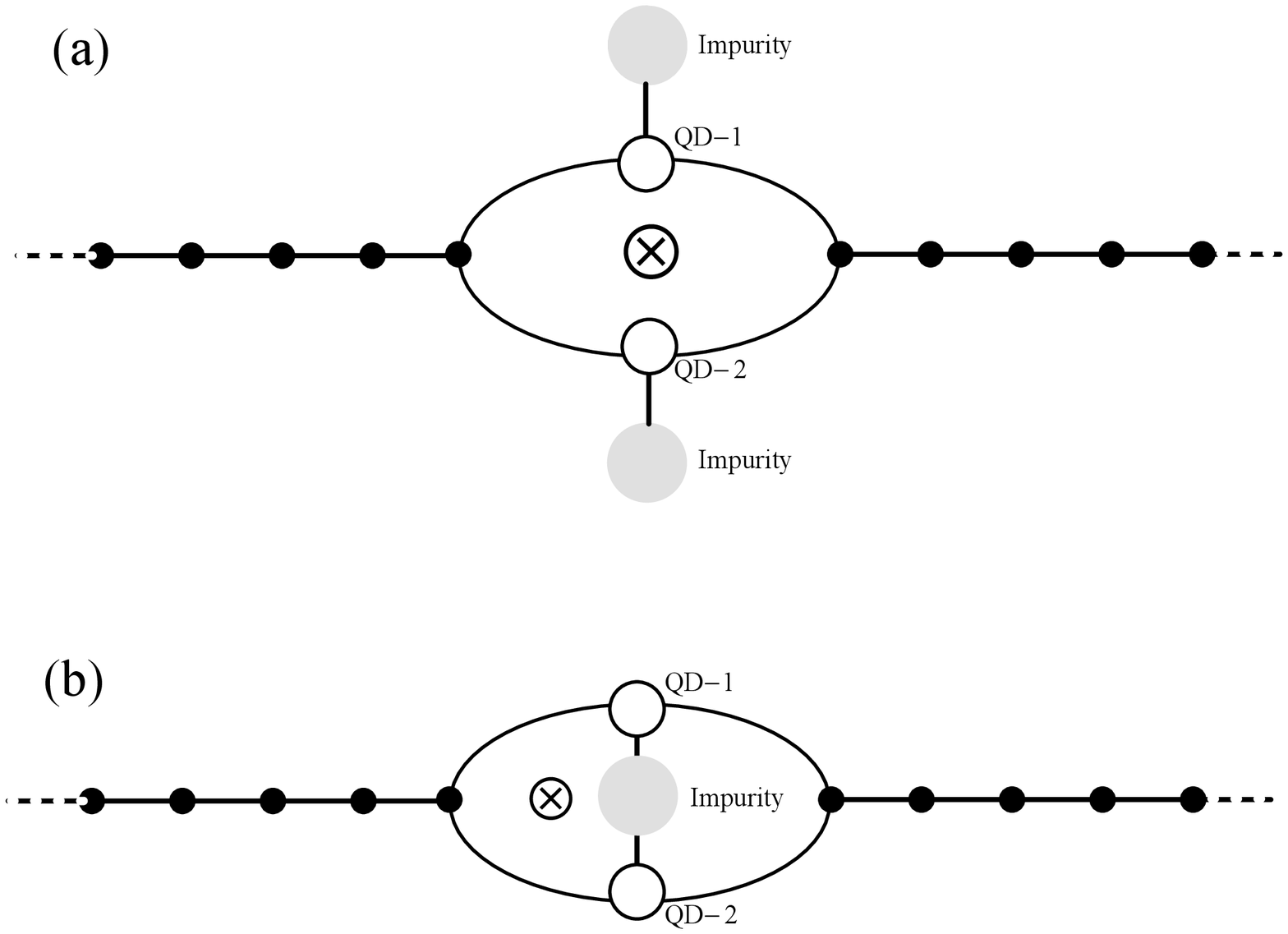}} \caption{ (a)
Schematic of a double QD AB interferometer with a local threading
magnetic flux by the presence of impurities. \label{structure}}

\end{figure}

The double QD AB interferometer under consideration is schematically
illustrated in Fig.\ref{structure}. The system Hamiltonian is
written as
\begin{eqnarray}
H_{0}&=&\sum_{\sigma k\alpha\in L,R}\varepsilon _{\alpha k}c_{\alpha
k\sigma}^\dag c_{\alpha k\sigma}+\sum_{\sigma,j=1}^{2}\varepsilon
_{j}d_{j\sigma}^\dag
d_{j\sigma}\notag\\&&+\underset{\sigma,\alpha,k,j }{\sum }
V_{j\alpha,\sigma}d_{j\sigma}^\dag c_{\alpha k\sigma}+\mathrm
{H.c.}, \label{4}
\end{eqnarray}
where $c_{\alpha k\sigma}^\dag$ and $d_{j\sigma}^\dag$ ( $c_{\alpha
k\sigma}$ and $d_{j\sigma}$) with the spin
$\sigma=\uparrow,\downarrow$ is an operator to create (annihilate)
an electron of the state $|k,\sigma\rangle$ in lead-$\alpha$
($\alpha=L, R$) and $|j,\sigma\rangle$ in QD-$j$ ($j=1,2$).
$\varepsilon _{\alpha k}$ and $\varepsilon_j$ are the corresponding
single-particle energy. $V_{j\alpha,\sigma}$ denotes the QD-lead
coupling strength. We adopt a uniform QD-lead coupling configuration
which gives that
$V_{1\sss{L},\sigma}=V_{1\sss{R},\sigma}^*=Ve^{i(\phi/4+\sigma\varphi)}$
and $V_{2\sss{R},\sigma}=V_{2\sss{L},\sigma}^*=Ve^{i\phi/4}$ with
the constant $V$. The phase shift $\phi$ is associated with the
threaded magnetic flux $\Phi$ by a relation
$\phi=2\pi\Phi/\Phi_{0}$, in which $\Phi_{0}=h/e$ is the flux
quantum. Besides, if the Rashba interaction is applied to either QD,
eg., QD-1, $V_{1\alpha, \sigma}$ will involve the spin-dependent
phase $\sigma\varphi$, which arises from the electron spin
precession induced by the Rashba SO coupling\cite{Sun,Chi}.
\par
To study the electronic transport properties, the linear conductance
of the noninteracting system at zero temperature is obtained by the
Landauer-B\"{u}ttiker formula\cite{refLandauer}
\begin{equation}
\mathcal {G}=\sum_\sigma{\mathcal
G}_\sigma=\frac{e^{2}}{h}\sum_\sigma
T_\sigma(\omega)|_{\omega=\varepsilon_F}.\label{conductance}
\end{equation}
$T_\sigma(\omega)$, the transmission function, in terms of Green
function takes the form as\cite{refMeir1} $T_\sigma(\omega)=\mathrm
{Tr}[\Gamma^L_\sigma G^r_\sigma(\omega)\Gamma^R_\sigma
G^a_\sigma(\omega)]$. $\Gamma^\alpha_\sigma$, a $2\times 2$ matrix,
is defined as $[\Gamma^{\alpha}]_{jj',\sigma}=2\pi
V_{j\alpha,\sigma}V_{j'\alpha,\sigma}^*\rho(\omega)$ describing the
coupling strength between the QDs and lead-$\alpha$. We will ignore
the $\omega$-dependence of $\Gamma^{\alpha}_{jj',\sigma}$ since the
electron density of states in lead-$\alpha$, $\rho(\omega)$, can be
usually viewed as a constant. The retarded and advanced Green
functions in Fourier space are involved, which can be solved by
means of the nonequilibrium Green function technique and
equation-of-motion method.\cite{refMeir1,Gong1} By a straightforward
derivation, we obtain the retarded Green functions written in a
matrix form as
\begin{eqnarray}
G^r_\sigma(\omega)=\left[\begin{array}{cc} g_{1\sigma}(z)^{-1} & i\Gamma_{12,\sigma}\\
  i\Gamma_{21,\sigma}& g_{2\sigma}(z)^{-1}
\end{array}\right]^{-1}\ \label{green}.
\end{eqnarray}
$g_{j\sigma}(z)=(z-\varepsilon_{j}+i\Gamma_{jj,\sigma})^{-1}$ with
$z=\omega+i0^+$ and
$\Gamma_{jj',\sigma}=\frac{1}{2}([\Gamma^L]_{jj',\sigma}+[\Gamma^R]_{jj',\sigma})$.
In addition, the advanced Green function can be readily obtained via
a relation $[G^a_\sigma(\omega)]=[G^r_\sigma(\omega)]^\dag$. Based
on the above derivations, the analytical expression of the linear
conductance of $\sigma$-spin electron can be written out by taking
$\Gamma=2\pi V^2\rho(\omega)$, i.e.,
\begin{widetext}
\begin{equation}
{\mathcal G}_\sigma=\frac{e^2}{h}\frac{\varepsilon^2_{1}
+\varepsilon^2_{2}
+2\varepsilon_{1}\varepsilon_{2}\cos(\phi+2\sigma\varphi)}
{[\varepsilon_{1}\varepsilon_{2}-\Gamma^2+\Gamma^2\cos^2({\phi\over
2}+\sigma\varphi)]^2
+(\varepsilon_{1}+\varepsilon_{2})^2\Gamma^2}\Gamma^2 \label{basic2}
\end{equation}
\end{widetext}

\section{Numerical results and discussion \label{result}}
With the formulation developed in the section above, we can carry
out the numerical calculation to investigate the linear conductance
of this double-QD structure modulated by impurities coupled to QDs.
Prior to calculation, we take the strength of the uniform QD-lead
couplings $\Gamma$ as the unit of energy and the Fermi level
$\varepsilon_F$ as the zero point of energy.
\par
To begin with, we do not take the Rashba SO coupling into account (
then, the electron transport is independent of the spin index ), and
for simplicity we focus on the structure of uniform QD levels by
letting $\varepsilon_j=\varepsilon$. Thereby, in the absence of
impurity the electron transport results of this structure are
ascertained with the help of Eq.(\ref{basic2}): In the
zero-external-flux case ${\mathcal G}$ has a compact expression as
$\frac{2e^2}{h}\frac{4\Gamma^2} {\varepsilon^2+4\Gamma^2}$, the same
as the result of the single-channel electron transmission with the
QD level $\varepsilon$ and QD-lead coupling $2\Gamma$, whose value
is obviously determined by the relative values of QD levels with
respect to the Fermi level of this system; however, irrelevant to
the shift of QD levels, a threaded magnetic flux with the phase
factor $\phi=\pi$ can lead to the conductance equal to
zero.\cite{Konig}

\subsection{Local impurity-QD coupling }
For the case where in the interferometer arms there exist impurities
coupled to the QDs respectively, i.e., the impurities couple to QDs
locally, the Hamiltonian of this system should be rewritten as
$H=H_0+H_{I}+H_{T}$. $H_{I}$ describes electrons in the impurities,
which takes a form as
\begin{eqnarray}
H_{I}&=&\sum_{m\sigma} (\epsilon_{m}a_{m\sigma}^\dag
a_{m\sigma}+\chi_ma_{m+1,\sigma}^\dag a_{m\sigma}+\rm
{H.c.})\nonumber\\&&+\sum_{n\sigma} (e_{n}b_{n\sigma}^\dag
b_{n\sigma}+\lambda_nb_{n+1,\sigma}^\dag b_{n\sigma}+\rm
{H.c.}),\label{3}
\end{eqnarray}
with $a^{\dag}_{m\sigma}$ and $b^{\dag}_{n\sigma}$ ($a_{m\sigma}$
and $b_{n\sigma}$) ($m$, $n \geq 1$) denoting the creation
(annihilation) operator of electron in impurities. $\epsilon_m$ and
$e_n$ are the single-energy levels of the corresponding impurities,
whereas $\chi_m$ and $\lambda_n$ represent the coupling coefficients
between two impurities. $H_{T}$ describes the electron tunneling
between the QDs and impurities
\begin{equation}
H_{T} =\sum_{\sigma}( t_{1}a_{1\sigma}^\dag
d_{1\sigma}+t_{2}b_{1\sigma}^\dag d_{2\sigma}+{\mathrm {H.c.}}) ,
\label{4}
\end{equation}
in which $t_j$ are the couplings between QDs and impurities.
Consequently, in such a case we have
$g_{j\sigma}(z)=(z-\varepsilon_{j}-\Sigma_{j\sigma}+i\Gamma_{jj,\sigma})^{-1}$.
The selfenergies $\Sigma_{j\sigma}$, originating from the couplings
of impurities to the QDs of the interferometer, can be written out
explicitly as
$\Sigma_{1\sigma}=\frac{t_1^2}{z-\epsilon_1-\frac{\chi^2_1}{z-\epsilon_2-\cdots}}$
and
$\Sigma_{2\sigma}=\frac{t_2^2}{z-e_1-\frac{\lambda^2_1}{z-e_2-\cdots}}$\cite{refLiu}.
Therefore, the effect of the local impurities only consists in
renormalizing the QD levels as
$\tilde{\varepsilon}_{j\sigma}=\varepsilon_{j}+\Sigma_{j\sigma}$.

\begin{figure}
\centering \scalebox{0.35}{\includegraphics{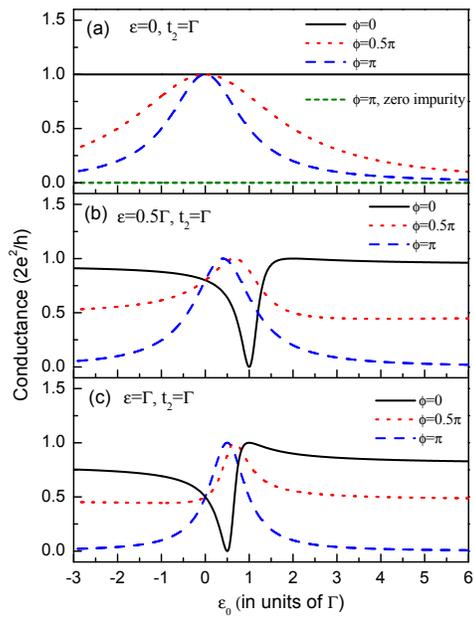}} \caption{ The
conductance spectra affected by the single-level impurity coupled to
QD-2. The uniform QD levels are assumed to be $0$ in (a),
$0.5\Gamma$ in (b), and $\Gamma$ in (c). \label{one-Im}}
\end{figure}

Then, we concentrate on the characteristics of the linear
conductance for the case of a single-level impurity coupled to
either QD of the interferometer (e.g., QD-2 here). By taking
$e_1=\varepsilon_0$, $t_2=\Gamma$, and $t_1=\lambda_1=0$, we show
the calculated results of the conductance $\cal G$ influenced by the
single-level impurity in Fig.\ref{one-Im}. From Fig.\ref{one-Im}(a),
we can find that when the QD levels are consistent with the Fermi
level, the conductance is equal to $2e^2/h$, the same as the
zero-impurity results. This means that in such a case the impurity
cannot affect the quantum transport through the structure. However,
a presented magnetic flux changes the role of impurity in this
quantum coherent transport process, namely, for the cases of a
finite magnetic flux through the interferometer, the profiles of
conductance \textit{vs} $\varepsilon_0$ present Breit-Wigner
lineshapes, and also, with the increase of magnetic flux the width
of the conductance peak becomes narrow, respectively corresponding
to the dotted and dashed lines in Fig.\ref{one-Im}(a). It is readily
found that in the presence of one impurity coupled to QD-2,
$\tilde{\varepsilon}_{2\sigma}$ is expressed as
$\varepsilon-t_2^2/\varepsilon_0$, so when $\varepsilon=0$ the
conductance can be simplified as ${\cal
G}=\frac{2e^2}{h}\frac{t^2_2}{\varepsilon_0^2(1-\cos^2{\phi\over
2})^2+t^2_2}$, and it is not difficult to understand the
contribution of impurity to the electronic transport under such a
condition.
\par
We next focus on the situations of the QD levels separated from the
Fermi level of the system (i.e., $\varepsilon\neq 0$), with the
results of $\varepsilon=0.5\Gamma$ shown in Fig.\ref{one-Im}(b).
From the figure, one can find that in the absence of magnetic flux,
the conductance spectrum shows a Fano lineshape with its
antiresonant point around the point of $\varepsilon_0=\Gamma$ and
the Fano resonance peak in the vicinity of $\varepsilon_0=2\Gamma$.
So, opposite to the $\varepsilon=0$ results the effect of impurity
here on the quantum transport is more apparent because of the
suppression or enhancement of the electron transmission with the
change of impurity level. When a magnetic flux is presented with
$\phi={\pi\over 2}$ the conductance profile does not show the
Breit-Wigner lineshape any more. Only with the further increase of
magnetic flux to $\phi=\pi$, the conductance curve, analogous to the
results of $\varepsilon=0$, also exhibits a Breit-Wigner lineshape,
but its peak departs from the zero point of energy.
Fig.\ref{one-Im}(c) offers us the conductance results of
$\varepsilon=\Gamma$, and, it can be seen that different from that
in Fig.\ref{one-Im}(b), when $\phi=0$ the Fano lineshape in the
conductance spectrum turns to be more clear, and its antiresonance
valley emerges in the vicinity of $\varepsilon_0={\Gamma\over 2}$
with the resonance peak at the point of $\varepsilon_0=\Gamma$.
Alternatively, although in the case of $\phi=\pi$ the lineshape of
conductance curve keeps `Breit-Wigner', its peakwidth is
correspondingly narrow and the conductance peak undergoes a futher
right shift. So far, the contribution of the single-level impurity
to the electron transport can be concluded as follows. At the
zero-external-field situation, only when the QD levels depart from
the Fermi level the impurity can modulate (enhance or weaken) the
electron traveling through the structure, since the appearance of
Fano lineshape in the conductance spectrum; on the other hand, when
there is a threaded magnetic flux with $\phi=\pi$, not only the
original destructive quantum interference can be changed, but also
the resonant tunneling can be realized with the appearance of the
Breit-Wigner lineshape in the conductance spectrum, irrelevant to
the change of QD levels.
\par
\begin{figure}
\centering \scalebox{0.35}{\includegraphics{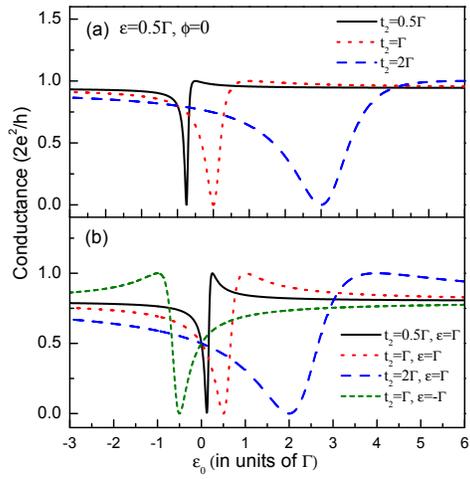}}\caption{ The
spectra of linear conductance with the change of the coupling
between QD-2 the impurity. \label{one-Im2}}
\end{figure}

With the help of the results in Fig. \ref{one-Im2}, this
single-level-impurity induced Fano effect can be detailedly
analyzed. By virtue of the results in Fig. \ref{one-Im2}(a) where
$\varepsilon=0.5\Gamma$, we can readily find that with the
strengthening of the impurity-QD coupling the Fano lineshape in the
conductance curve shifts right, followed by the wideness of
antiresonance valley proportionally. Via a further observation, we
know that the antiresonance position is tightly associated with the
impurity-QD coupling strength, and at the point of
$\varepsilon_0\approx t_2^2$ there occurs Fano antiresonance in
electron tunneling through such a system. That is to say, in this
configuration the antiresonant point in the conductance spectrum can
reflect the coupling strength between the impurity and QD, which may
be helpful for the current experiment in the low-dimensional
physics. Then, by fixing the QD levels at $\varepsilon=\Gamma$, we
plot the conductance spectra with the increase of $t_2$, as shown in
Fig.\ref{one-Im2}(b). Distinctly, although the antiresonance point
shifts right with the increment of the QD-impurity coupling, the
relation between the antiresonance and the QD-impurity coupling is
different from the results in the above case, i.e., here the
antiresonant point comes into being at the position of
$\varepsilon_0\sim {t_2^2\over 2}$ in principle. Therefore, it is
clear that the presentation of the antiresonance is dependent on
both the QD-impurity coupling strength as well as the separation of
the QD levels from the Fermi surface.

\par
In order to discuss the impurity-induced Fano resonances we would
like to write the conductance expression as its Fano form. Without
loss of generality, the Fano form of the conductance formula can be
easily obtained
\begin{equation}
\mathcal
G=\sum_\sigma\frac{e^2}{h}T_{b\sigma}\frac{|e_\sigma+q_\sigma|^2}{e^2_\sigma+1},
\label{fanoform}
\end{equation}
where
$T_{b\sigma}=\Gamma^L_{11,\sigma}\Gamma^R_{11,\sigma}/[\tilde{\varepsilon}_{1\sigma}^2+\Gamma^2_{11,\sigma}]$,
$e_\sigma=-\mathrm {Re}G^r_{22,\sigma}/\mathrm {Im}G^r_{22,\sigma}$,
and
$q_\sigma=-\frac{\tilde{\varepsilon}_{1\sigma}}{\Gamma_{11,\sigma}}
(\Gamma^L_{12,\sigma}\Gamma^R_{21,\sigma}\Gamma_{11,\sigma}-
T_{b\sigma}\Gamma_{12,\sigma}\Gamma_{21,\sigma}\Gamma_{11,\sigma})/({\Gamma^L_{11,\sigma}
\Gamma^R_{11,\sigma}\Gamma_{22,\sigma}-T_{b\sigma}\Gamma_{12,\sigma}\Gamma_{21,\sigma}\Gamma_{11,\sigma}})$.
Under the condition of uniform QD-lead coupling, $q_\sigma$ can be
simplified as
$q_\sigma=-\frac{\tilde{\varepsilon}_{1\sigma}}{\Gamma}[e^{i(\phi+2\sigma\varphi)}-T_{b\sigma}
\cos^2(\frac{\phi}{2}+\sigma\varphi)]/[1-T_{b\sigma}
\cos^2(\frac{\phi}{2}+\sigma\varphi)]$ and
$e_\sigma=-\frac{\tilde{\varepsilon}_{2\sigma}}{\Gamma}[1+T_{b\sigma}\cos^2(\frac{\phi}{2}+\sigma\varphi)\frac{
\varepsilon_{1\sigma}}{\tilde{\varepsilon}_{2\sigma}}]/[1-T_{b\sigma}
\cos^2(\frac{\phi}{2}+\sigma\varphi)]$. It is seen that, for the
case shown in Fig.\ref{one-Im}(a) in which the Rashba interaction is
absent and the uniform QD levels coincide with the Fermi level,
$q_\sigma$ is equal to zero, so the conductance profile does not
exhibit the Fano lineshape. Concretely, when $\phi=0$ $e_\sigma$
gets close to infinity all long, which induces the result of ${\cal
G}\equiv2e^2/h$. With regard to the results shown in
Fig.\ref{one-Im}(b)-(c), in the zero-magnetic-field case the
separation of $\varepsilon$ from the Fermi level makes $q_\sigma$
nonzero and $e_{\sigma}$ convergent, thereby, the Fano lineshape
comes about in the conductance spectrum. Moreover, the antiresonant
points in the conductance spectra can be ascertained by considering
$e_\sigma+q_\sigma=0$. Via a simple derivation, the position of
antiresonance can be obtained analytically, when
$\varepsilon_0={t_2^2\over2\varepsilon}$ the conductance encounters
zero which means the occurrence of the antiresonance effect. Based
on this result, one can understand the quantitative dependence of
the antiresonance point on both the strength of impurity-QD coupling
and the values of QD levels. Therefore, in such a structure, the
occurrence of Fano antiresonance remarkably differs from those in
the single-channel system with impurity, where the antiresonance
occurs once the impurity level is aligned with the Fermi
level\cite{T-shaped}. The position of Fano peak can be obtained as
well, accordingly, just at the point of $t_2^2\over\varepsilon$
there emerges the Fano resonance. Meanwhile, it is certain that
changing the position of QD levels with respect to the Fermi level
can lead to the variation of the sign of $\varepsilon$, then, the
sign of $q_\sigma$ alters which brings out the inversion of the Fano
lineshape, corresponding to the shot-dashed line in
Fig.\ref{one-Im2}. Up to now, in this structure the Fano effect
driven by a single-level impurity has been clarified.

\par
According to the results in Fig.\ref{one-Im}, when $\phi=\pi$
$q_\sigma=\frac{\tilde{\varepsilon}_{1\sigma}}{\Gamma}$ is real, but
the conductance curves present Breit-Wigner lineshapes, independent
of the deviation of QD levels from the Fermi level. The only
phenomenon is that with the increase of $\varepsilon$, the
conductance peak becomes narrow following its right shift. Thus,
such results do not coincide with the previous discussions that only
an imaginary $q_\sigma$ could cause the appearance of a Breit-Wigner
lineshape in the corresponding conductance.\cite{gongpe} We have to
continue to study the behaviors of $e_\sigma+q_\sigma$ when
$\phi=\pi$. Then, we find that in such a case
$e_\sigma+q_\sigma={t_2^2\over \Gamma \varepsilon_0}$ has no
opportunity to be zero, irrelevant to the QD levels in the quantum
transport regime. By a simplification, the conductance can be
written explicitly as ${\mathcal
G}=\frac{2e^2}{h}T_{b\sigma}t_2^4/[(\varepsilon_0-\frac{\varepsilon
t_2^2}{\varepsilon^2+\Gamma^2})^2
+\frac{\Gamma^2t_2^4}{\varepsilon^2+\Gamma^2}]$. Based on such a
result, the exhibition and properties of the Breit-Wigner lineshape
in the conductance curve can be well understood, i.e., decreasing
the QD-impurity coupling or increasing the value of QD levels can
lead to the narrowness of the conductance peak. Well, in the case of
$\varepsilon=\Gamma$, the quantities $\frac{\varepsilon
t_2^2}{\varepsilon^2+\Gamma^2}=\frac{ t_2^2}{2\varepsilon }$, which
leads to the superposition of the antiresonant point in the $\phi=0$
conductance spectrum and the conductance peak of $\phi=\pi$. In
addition, all the results above tell us that the magnetic-induced
inversion of Fano lineshapes is unfeasible in this model.
\par
\begin{figure}
\centering \scalebox{0.35}{\includegraphics{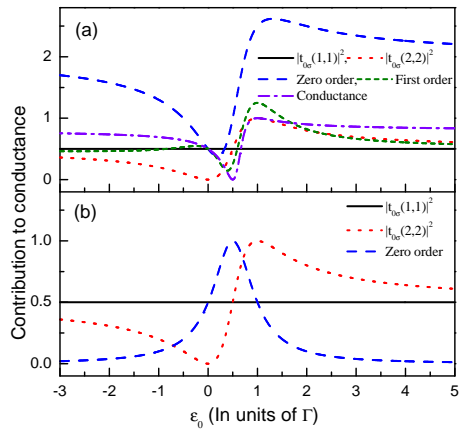}}\caption{ The
Feynman-path analysis of the impurity-modified quantum interference.
\label{path}}
\end{figure}

Although the above analysis is helpful for understanding the
properties of the impurity-driven electron transport, one has to
know that the underlying physics in the electron motion of such a
structure should be quantum interference, which can be described by
means of the language of Feynman path\cite{gongpe}. To illustrate
this issue, we rewrite the electron transmission function as
$T_\sigma(\omega)=\mathrm {Tr}[\Gamma^L_\sigma
G^r_\sigma\Gamma^R_\sigma G^a_\sigma]=|\sum\limits_{j,
l=1}^2t_\sigma(j,l)|^2$, where the electron transmission
coefficients are defined as
$t(j,l)=\bar{V}_{\sss{L}j,\sigma}G^r_{jl,\sigma}\bar{V}_{l\sss{R},\sigma}$
with $\bar{V}_{j\alpha,\sigma}=\bar{V}_{\alpha
j,\sigma}^*=V_{j\alpha,\sigma}\sqrt{2\pi\rho(\omega)}$. Besides, it
is necessary to define
$\widetilde{V}_{j\alpha,\sigma}=\widetilde{V}_{\alpha
j,\sigma}^*=V_{j\alpha,\sigma}\sqrt{\pi\rho(\omega)}$ for the
following discussion. At the beginning, we expand the Green function
into an infinite geometric series. Taking $G_{11}^r$ as an example,
we have
$G^r_{11,\sigma}=g^{-1}_{2\sigma}/[g^{-1}_{1\sigma}g^{-1}_{2\sigma}+\Gamma_{12,\sigma}\Gamma_{21,\sigma}]
=\sum\limits_{j=0}^\infty
g_{1\sigma}(-g_{1\sigma}g_{2\sigma}\Gamma_{12,\sigma}\Gamma_{21,\sigma})^j$.
Thereby, we can express the transmission coefficient $t_\sigma(1,1)$
as a summation of Feynman paths with different orders, i.e.,
$t_\sigma(1,1)=\sum\limits_{j=0}^\infty\bar{V}_{\sss{L}1,\sigma}
g_{1\sigma}(-g_{1\sigma}g_{2\sigma}\Gamma_{12,\sigma}\Gamma_{21,\sigma})^j\bar{V}_{1\sss{R},\sigma}=\sum\limits_{j=0}^\infty
t_{j\sigma}(1,1)$. By the same token, we can expand other
transmission coefficients as a summation of Feynman paths:
$t_{\sigma}(2,2)=\sum\limits_{j=0}^\infty
\bar{V}_{\sss{L}2,\sigma}g_{2\sigma}(-g_{2\sigma}g_{1\sigma}\Gamma_{12,\sigma}\Gamma_{21,\sigma})^j\bar{V}_{2\sss{R},\sigma}
=\sum\limits_{j=0}^\infty t_{j\sigma}(2,2)$,
$t_{\sigma}(1,2)=\sum\limits_{j=1}^\infty
i\bar{V}_{\sss{L}1,\sigma}(-g_{1\sigma}g_{2\sigma}\Gamma_{12,\sigma})^{j}\Gamma_{21,\sigma}^{j-1}\bar{V}_{2\sss{R},\sigma}
=\sum\limits_{j=1}^\infty t_{j\sigma}(1,2)$, and
$t_{\sigma}(2,1)=\sum\limits_{j=1}^\infty
i\bar{V}_{\sss{L}2,\sigma}(-g_{1\sigma}g_{2\sigma}\Gamma_{21,\sigma})^{j}\Gamma^{j-1}_{12,\sigma}\bar{V}_{1\sss{R},\sigma}
=\sum\limits_{j=1}^\infty t_{j\sigma}(2,1)$. With the above
analysis, we find that quantum interference occurs among infinite
Feynman paths, and only for the case of $\phi=\pi$ all the
higher-order paths vanish because of the destructive interference
among them and the two zero-order paths[$t_{0\sigma}(1,1)$ and
$t_{0\sigma}(2,2)$] remain. We then analyze the respective
contributions of the leading Feynman paths to the quantum
interference, as shown in Fig.\ref{path}. From the figure, it is
found that $t_{0\sigma}(1,1)$ provides a nonresonant path for the
electron transmission. But, in the other interferometer arm, the
coupling between the impurity and QD-2 modulates the feature of
$|t_{0\sigma}(2,2)|^2$ nontrivially because it presents an
antisymmetric lineshape with the shift of $\varepsilon_0$. Of
course, the phase of $t_{0\sigma}(2,2)$ can also be tuned with the
shift of the impurity level. As a result, the interference between
these two paths induces the antisymmetric lineshape of the
contribution to the conductance in spite of its being greater than
unity. When the contribution of the first-order paths is taken into
account, the contribution of low-order paths to conductance gets
close to the curve of the standard conductance. With this, the
occurrence of Fano interference in this system can be well
clarified, i.e., just the existence of impurity modifies the phase
of electron waves through QD-2 of the interferometer. On the other
hand, in the presence of magnetic flux $\phi=\pi$, the high-order
modification can be safely ignored due to the destructive
interference of high-order Feynman paths, and the interference
between the two paths gives rise to the appearance of Breit-Wigner
lineshape in the conductance spectrum, as shown in
Fig.\ref{path}(b).

\par
\begin{figure}
\centering \scalebox{0.35}{\includegraphics{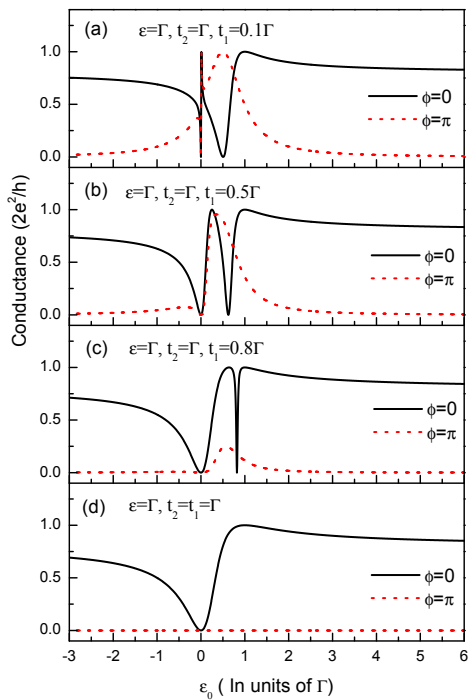}}\caption{ The
conductance curves in the case where each QD couples to a
single-level impurity. \label{twoIm}}
\end{figure}

With regard to the situation where in both arms there exist
impurities, we first would like to consider the simple case that
there is a single-level impurity respectively side-coupled to each
QD of the interferometer. The corresponding results are shown in
Fig.\ref{twoIm} in which $\epsilon_1=e_1=\varepsilon_0$ and
$\chi_1=\lambda_1=0$. Here, by fixing $t_2$ at $\Gamma$ and
increasing the amplitude of $t_1$, the interplay between the
impurities in the two arms on the quantum interference is easy to be
understood. Just in Fig.\ref{twoIm}(a), we can see that the nonzero
$t_1$ changes the electronic transport strikingly with its
complicating the original Fano interference, even if a weak coupling
between the impurity and QD-1 is considered. To be concrete, in the
case of zero magnetic flux, except the antiresonance point at the
position of $\varepsilon_0={t_2^2\over 2\varepsilon}$ the other
antiresonance occurs in the vicinity of $\varepsilon_0=0$. By virtue
of the results in previous works, this phenomenon can be explained
as follows. In the presence of an impurity side-coupled to one QD,
the electron transport in the corresponding channel will be
suppressed completely when the impurity is aligned with the Fermi
level\cite{T-shaped}. Thus, when both the impurity levels are
identical with the Fermi level, the electron transmission in both
channels of the interferometer are restraint, which, independent of
the application of magnetic flux, results in the appearance of
antiresonance at the point of $\varepsilon_0=0$. So, when
$\phi=\pi$, there is also a Fano antiresonance in the conductance
spectrum at the point of $\varepsilon_0=0$. In addition, with the
increment of $t_1$ from $0.2\Gamma$ to $\Gamma$, the original Fano
antiresonance ( at the point of $\varepsilon_0={t_2^2\over
2\varepsilon}$ ) shifts right and the antiresonance valley becomes
narrow until its disappearance, but, the antiresonant point at the
Fermi level is fixed and the corresponding antiresonance valley
widens. This is for reason that the increase of $t_1$ the property
of $g_{1\sigma}$ gets close to that of $g_{2\sigma}$, and, just when
$t_1=t_2$ $g_{1\sigma}$ has the same properties as $g_{2\sigma}$,
which drives the present electron transport similar to that in the
single-channel case (i.e., the T-shaped double QDs ). Then, one can
readily understand the emergence of conductance zero at the position
of $\varepsilon=0$ in the zero-magnetic-flux case, as shown in
Fig.\ref{twoIm}(d).

\par
\begin{figure}
\centering \scalebox{0.35}{\includegraphics{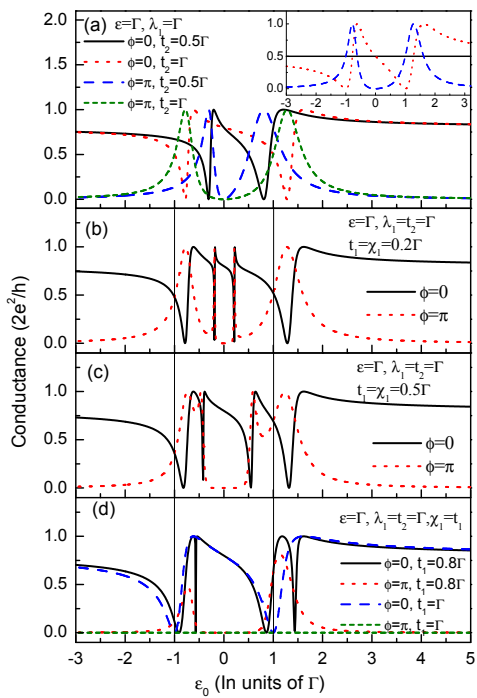}}\caption{(a)
The conductance spectra in the presence of a double-level impurity
side-coupled to QD-2 with the zero-order paths in the inset. (b)-(d)
The conductance curves in the case where each QD couples to a
double-level impurity. \label{twoIm2}}
\end{figure}

There is no doubt that the consideration of multi-level impurities
can further influence the quantum interference in this structure. As
shown in Fig.\ref{twoIm2}, we plot the conductance spectra in the
case of double-level impurities considered. We can find, from
Fig.\ref{twoIm2}(a) with $\varepsilon=\Gamma$ and $t_1=\lambda_2=0$,
that when only one double-level impurity is presented in either arm
of the interferometer, in the case of zero magnetic flux there are
two Fano lineshape existing in the conductance spectrum, the space
of which is consistent with the impurity-level space. Similar to
those in the case of single-level impurity, the Fano antiresonant
points in conductance spectrum of $\phi=0$ corresponding to the
conductance peak in the case of $\phi=\pi$. But, here between the
$\phi=\pi$ conductance peaks an antiresonance comes into being,
which can be clarified by paying attention to the interference
between $t_{0\sigma}(1,1)$ and $t_{0\sigma}(2,2)$, described by the
inset of Fig.\ref{twoIm2}(a). For the case where in each arm there
is a double-level impurity, the lineshapes of the conductance become
more complex, as shown in Fig.\ref{twoIm2}(b)-(d). In such a case,
we fix the values of $t_2$ and $\lambda_1$, while change the
amplitudes of $t_1$ and $\chi_1$ to analyze the variation of
conductance by the two double-level impurities. Surely, both the
impurities in different channels contribute to the antiresonances in
the $\phi=0$ conductance curve and resonances in $\phi=\pi$
conductance profiles, respectively. It is seen that when $\phi=\pi$,
compared with the results of one impurity, the antiresonance in the
vicinity of Fermi level becomes an insulating band and it is more
clear with the increase of $t_1$, which also arises from the similar
properties of $g_{1\sigma}$ and $g_{2\sigma}$. Similar to the
results in Fig.\ref{twoIm}(d), when the identical impurities
respectively couple to the QDs with uniform couplings, the Fano
antiresonance is only relevant to the eignlevels of the impurities (
$\varepsilon_0\pm\lambda_1$ here ).

\par
\begin{figure}
\centering \scalebox{0.35}{\includegraphics{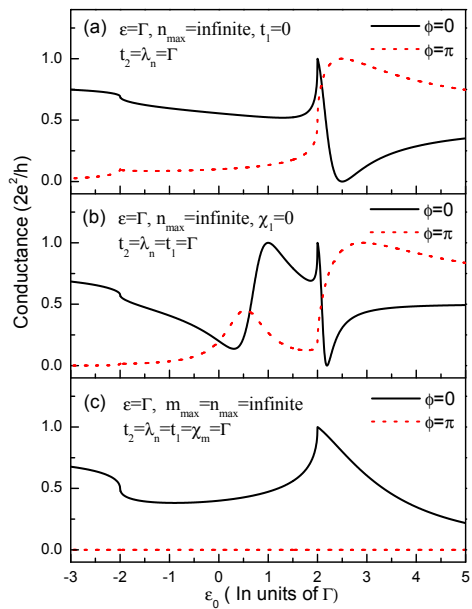}}\caption{The
linear conductance spectra for the case of infinite-level impurities
side-coupled to the QDs of the interferometer. \label{continuum}}
\end{figure}

As for the extreme case where the levels of impurities become
continuous, by assuming $\epsilon_m=e_n=\varepsilon_0$ and
$t_j=\chi_m=\lambda_n=\Gamma$ we can obtain
$\Sigma_{1(2)\sigma}=\frac{1}{2}(-\varepsilon_0-i\sqrt{4\Gamma^2-\varepsilon_0^2})$
in such a case. Under the condition of $|\varepsilon_0|<2\Gamma$,
${\mathrm {Im}}\Sigma_{1(2)\sigma}\neq0$. So it can be anticipated
that in electron transport through the QDs, there will occur the
notable inelastic scattering. As shown in Fig.\ref{continuum}(a),
even if only there exist continuous-level impurity in one arm of the
interferometer, the electron coherent tunneling through the
corresponding channel is seriously restraint. Only at the upper edge
of the energy band the Fano interference emerges in the absence of
magnetic flux, which is due to that ${\mathrm
{Im}}\Sigma_{1(2)\sigma}\rightarrow 0$ in this region and then the
role of impurity is similar to that of single-level one. However, in
Fig.\ref{continuum}(b) it is seen that the appearance of an
additional concrete-level impurity in the other interferometer arm
can enhance the electronic transport. When there exist
continuous-level impurities coupled to both the QDs, no Fano
lineshape comes about in the conductance spectrum any more.

\par
\begin{figure}
\centering \scalebox{0.35}{\includegraphics{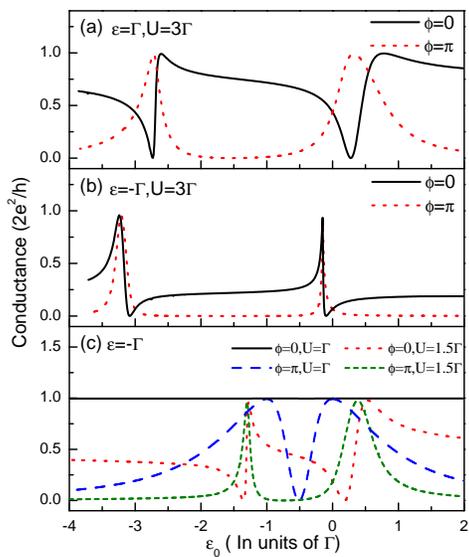}}\caption{The
impurity-related linear conductance spectra in the case of the
many-body terms being considered with $U=3\Gamma$ in (a) and (b). In
(c), $U=\Gamma$ and $1.5\Gamma$, respectively. \label{manybody}}
\end{figure}

Next, we have to make a remark regarding the many-body effect which
we have by far ignored. As is known, the many-body effect is an
important origin for the peculiar transport properties in QDs.
Usually, the many-body effect is incorporated by considering only
the intradot Coulomb repulsion, i.e., the Hubbard term. If the
Hubbard interaction is not very strong, we can truncate the
equations of motion of the Green functions to the second order. By a
straightforward derivation, we find that in such an approximation,
only the relevant QD and impurity levels are renormalized, i.e.,
$\varepsilon_{j\sigma}=
\varepsilon_j(\frac{z-\varepsilon_j-U_j}{z-\varepsilon_j-U_j+U_j\langle
n_{j\bar{\sigma}}\rangle})$, $\epsilon_{m\sigma}=
\epsilon_m(\frac{z-\epsilon_m-U_m}{z-\epsilon_m-U_m+U_m\langle
n_{m\bar{\sigma}}\rangle})$, and $e_{n\sigma}=
e_n(\frac{z-e_n-U_n}{z-e_n-U_n+U_n\langle
n_{n\bar{\sigma}}\rangle})$, correspondingly, the formula
Eq.(\ref{conductance}) is still feasible to calculate the linear
conductance.\cite{refMeir1,refLiu} As a typical case, in
Fig.\ref{manybody}, by taking the many-body terms into account, we
investigate the linear conductance modified by a single-level
impurity, with the structure parameters the same as those in
Fig.\ref{one-Im}(c). From the figure, we can find that when the
uniform many-body strength is considered, the conductance spectra
are divided into two groups, since the energy level $\varepsilon_j$
( $\epsilon_m$ or $e_n$ ) splits into two, i.e., $\varepsilon_j$ and
$\varepsilon_j+U$ ( $\epsilon_m$ and $\epsilon_m+U$, or $e_n$ and
$e_n+U$ ). It is seen that only in the case of $\varepsilon=\Gamma$
the conductance lineshape in each group keeps the electron transport
properties of the noninteracting case. But when
$\varepsilon=-\Gamma$, the effect of the electron interaction on the
electron transport is more remarkable and the lineshapes of the
conductance spectra in each group are tightly influenced by the
Coulomb strength, as shown in Fig.\ref{manybody}(b)-(c), which
presents as the disappearance or inversion of the Fano lineshapes
caused by the different-strength Coulomb interactions. Such results
are due to that the Coulomb repulsion brings about the change of
$q_\sigma$'s sign ( $+$ or $-$ ). When the appropriate electron
interaction in QD-1 is considered, the renormalized level
$\varepsilon_{1\sigma}$ is written as
$\varepsilon_1(\frac{z-\varepsilon_1-U_1}{z-\varepsilon_1-U_1+U_1\langle
n_{1\bar{\sigma}}\rangle})$, so, the signs of both
$\varepsilon_{1\sigma}$ and the Fano parameter $q_{\sigma}$ ( here
$q_{\sigma}=-\varepsilon_{1\sigma}/\Gamma$ ) is obviously dependent
on the Coulomb strength $U_1$.

\subsection{Nonlocal impurity-QD coupling}
With respect to the case of the impurity coupled to both QDs
nonlocally, the Hamiltonian $H_I$ and $H_T$ should be respectively
expressed as
\begin{eqnarray}
H_I=\sum_{\sigma, m} \epsilon_{m}a_{m\sigma}^\dag
a_{m\sigma}+\sum_{\sigma, m=1}^{N-1}\chi_ma_{m+1,\sigma}^\dag
a_{m\sigma}+\rm {H.c.}
\end{eqnarray}
and
\begin{eqnarray}
H_T=\sum_{\sigma}( t_{1}a_{1\sigma}^\dag
d_{1\sigma}+t_{2}a_{N\sigma}^\dag d_{2\sigma}+{\mathrm {H.c.}}).
\end{eqnarray}
For simplicity, by means of the theory in Ref.\onlinecite {jpcm} we
can work out the Green functions involved so as to the conductance.
\par
\begin{figure}
\centering \scalebox{0.35}{\includegraphics{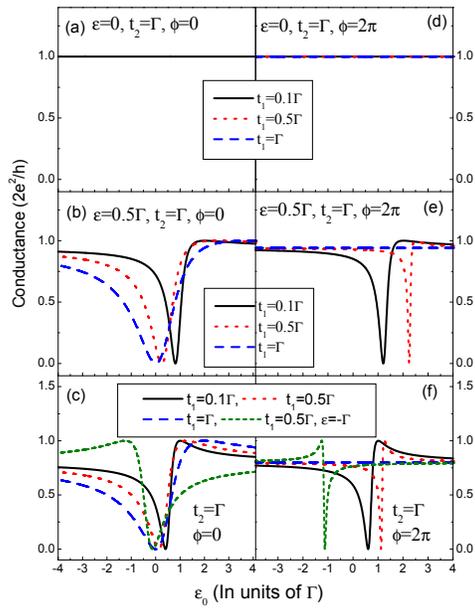}}\caption{ The
conductance curves in the presence of a nonlocal single-level
impurity coupled to the QDs, for the cases of $\phi=0$ and
$\phi=2\pi$. \label{both1}}
\end{figure}

Then, in Fig.\ref{both1} the situation of a single-level impurity is
first discussed, and by fixing $t_2$ at $\Gamma$ we investigate the
influence of the presence of $t_1$ on the conductance properties.
From this figure it can be found that when no magnetic flux is taken
into account, similar to the results in Fig.\ref{one-Im2}, only in
the cases of QD levels separated from the Fermi level there emerge
the Fano lineshapes in the corresponding conductance spectra and the
excess of QD levels to the Fermi level gives rise to the inversion
of the Fano lineshapes. However, the appearance of nonzero $t_1$
shifts the antiresonance position in the conductance curve to a
great extent. First, the antiresonance does not appear at the point
of $\varepsilon_0={t_2^2\over 2\varepsilon}$ any more but with the
increase of $t_1$ it gets close to the zero point of energy,
followed by the wideness of the antiresonance valley. Just when
$t_1=t_2=\Gamma$ the antiresonance occurs at the point of
$\varepsilon_0=0$, as shown in Fig.\ref{both1}(b)-(c). Besides, it
is of importance that the simultaneous couplings of the impurity to
QDs alter the period of conductance change ( \textit{vs} magnetic
flux ) from $2\pi$ to $4\pi$. Concretely, when the magnetic flux
increases to $\phi=2\pi$, the conductances present much differences
from those in the zero-magnetic-flux case, as shown in
Fig.\ref{both1}(e)-(f), i.e., in such a case the increase of $t_1$
induces the shift of antiresonance point to the high-energy
direction with the narrowness of the antiresonance valley, until the
disappearance of antiresonance when $t_1=\Gamma$ in which the
conductance becomes irrelevant to the tuning of $\varepsilon_0$.
\par
Analogous to the discussion in the above subsection, for convenience
of analyzing the Fano antiresonance effect here we also rewrite the
conductance expression into its Fano form. First, the retarded Green
functions that describe the electron motion can be written in a
matrix form as
\begin{eqnarray}
G^r_\sigma(\omega)=g_{\sigma}(z)^{-1}\cdot\left[\begin{array}{cc} \tilde{g}_{1\sigma}(z)^{-1}& i\tilde{\Gamma}_{12,\sigma}\\
  i\tilde{\Gamma}_{21,\sigma}&\tilde{g}_{2\sigma}(z)^{-1}
\end{array}\right]^{-1}\ \label{green}
\end{eqnarray}
where
$\tilde{g}_{j\sigma}(z)^{-1}=g_{j\sigma}(z)^{-1}g_{\sigma}(z)^{-1}-|t_{j}|^2$,
$\tilde{\Gamma}_{12,\sigma}=it_1t_2+\Gamma_{12,\sigma}g_{\sigma}(z)^{-1}$,
and
$\tilde{\Gamma}_{21,\sigma}=it^*_1t^*_2+\Gamma_{21,\sigma}g_{\sigma}(z)^{-1}$,
with $g_{\sigma}(z)=[z-\varepsilon_0]^{-1}$ being the impurity Green
function in the absence of impurity-QD couplings. Consequently, by
letting
$T_{b\sigma}=\Gamma^L_{11,\sigma}\Gamma^R_{11,\sigma}/[\tilde{\varepsilon}_{1}^2+\Gamma^2_{11,\sigma}]$
( with $\tilde{\varepsilon}_j=\varepsilon_j+|t_j|^2g_{\sigma}$) and
$e_\sigma=-\mathrm {Re}G^r_{22,\sigma}/\mathrm {Im}G^r_{22,\sigma}$
the Fano form of the conductance formula can be obtained
\begin{equation}
\mathcal
G=\sum_\sigma\frac{e^2}{h}T_{b\sigma}\frac{|e_\sigma+\tilde{q}_\sigma|^2}{e^2_\sigma+1},
\label{fanoform2}
\end{equation}
and the new Fano parameter is defined as
$\tilde{q}_\sigma=-\frac{\tilde{\varepsilon}_{1}}{\Gamma}[e^{i(\phi+2\sigma\varphi)}-{\cal
A}_{\sigma}T_{b\sigma}]/[1-T_{b\sigma}({\cal A}_{\sigma}+{\cal
B}_{\sigma}\tilde{\varepsilon}_{1})]$, with ${\cal
A}_{\sigma}=(\Gamma_{12,\sigma}\Gamma_{21,\sigma}-|t_1t_2|^2g_{\sigma}^2)/\Gamma^2$
and ${\cal
B}_{\sigma}=(t_1t_2\Gamma_{12,\sigma}+t^*_1t^*_2\Gamma_{21,\sigma})g_{\sigma}/\Gamma^3$.
Then, with the help of the results above, the Fano antiresonance
point can be clarified, i.e., at the point of
$\varepsilon_0={|t_1-t_2|^2\over 2\varepsilon}$ the conductance
becomes zero in the absence of magnetic flux. However, when
$\phi=2\pi$ the antiresonant point comes into being at the position
of $\varepsilon_0={|t_1+t_2|^2\over 2\varepsilon}$, except for the
case of $t_1=t_2$ where ${\cal G}\sim{4\Gamma^2\over
\varepsilon^2+4\Gamma^2}$ is extremely simplified and independent of
$\varepsilon_0$. On the other hand, notice that the quantum
interference in this structure is completely different from the
system in the above subsection, and it seems that the quantum
interference of this case is relatively complicated since the
nonlocal couplings of the impurity to QDs provides an additional
channel for the electron transmission. In order to analyze the
quantum interference in such a situation, one has to deal with this
model in the molecular-orbital representation, because the coupled
structure formed by the impurity and QDs can be viewed as a new QD
molecule, and just the quantum interference among its eigenstates
causes the present conductance results. The detailed demonstration
in Ref.\onlinecite{jpcm} can help us clarify the picture of quantum
interference here.
\par
\begin{figure}
\centering \scalebox{0.35}{\includegraphics{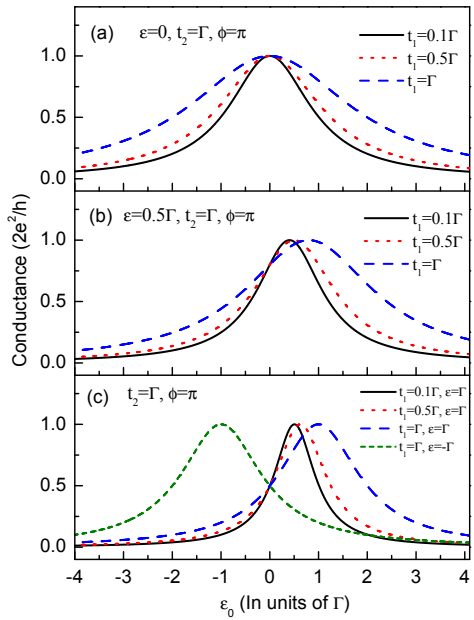}}\caption{The
conductance curves with a nonlocal single-level impurity coupled to
the QDs in the case of $\phi=\pi$. \label{both2}}
\end{figure}

We next show the conductance spectra in the case of the magnetic
flux phase factor $\phi=\pi$. It is clear that the conductance
spectra always exhibit Breit-Wigner lineshapes, independent of the
change of $t_1$ and $\varepsilon$. But, the effect of the nonzero
$t_1$ on the modulation of conductance lineshape can not be ignored.
As presented by Fig.\ref{both2}(a) when the QD levels are the same
as the Fermi level, i.e., $\varepsilon=0$, the increment of $t_1$
causes the wideness of the conductance peak. In addition, the
results in Fig.\ref{both2}(b)-(c) suggest that for the case of the
finite-value QD levels, with the increase of $t_1$ the conductance
peak shifts right as well, accompanied by its widening. When paying
attention to the Fano form of the conductance in
Eq.(\ref{fanoform2}), we find that there is no probability for
$e_\sigma+\tilde{q}_{\sigma}$ to be equal to zero, but the
conductance expression can be written as ${\mathcal
G}=\frac{2e^2}{h}T_{b\sigma}(t_1^2+t_2^2)^2/[(\varepsilon_0-\frac{\varepsilon
(t_1^2+t_2^2)}{\varepsilon^2+\Gamma^2})^2
+\frac{\Gamma^2(t_1^2+t_2^2)^2}{\varepsilon^2+\Gamma^2}]$, by which
the conductance properties in this case can be well understood.

\par
\begin{figure}
\centering \scalebox{0.35}{\includegraphics{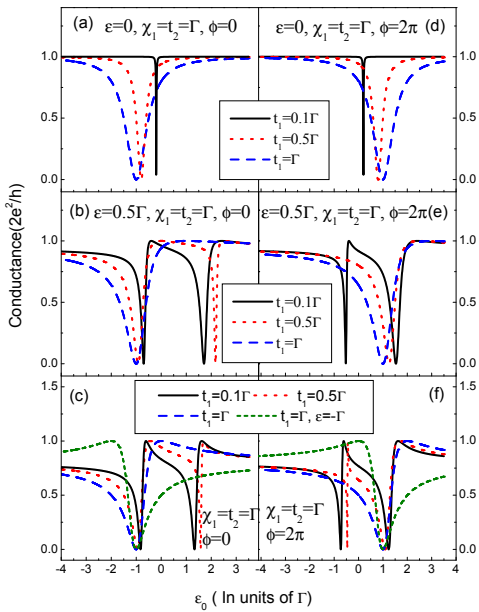}} \caption{The
conductance curves with a nonlocal double-level impurity coupled to
the QDs, for the cases of $\phi=0$ and $\phi=2\pi$. \label{both3}}
\end{figure}

When there is a nonlocal double-level impurity coupled to the QDs,
from Fig.\ref{both3} where the conductances of $\phi=0$ and
$\phi=2\pi$ are discussed, we can find that the role of the impurity
in changing the quantum transport is also remarkable in comparison
with the results of $t_1=0$ in Fig.6(a). Just as exhibited in
Fig.\ref{both3}(a) and (d), in this case even if $\varepsilon$ is
set to be zero, the extra coupling between the impurity and QD-1
leads to the occurrence of antiresonance: For the zero-magnetic-flux
case by the increase of $t_1$ to $t_1=\Gamma$ the antiresonance
position shifts left to $\varepsilon_0=-\Gamma$ but when $\phi=2\pi$
the antiresonant point shifts right to $\varepsilon_0=\Gamma$
following the wideness of the antiresonance valley. When the QD
levels are separated from the Fermi level with
$\varepsilon=0.5\Gamma$ in Fig.\ref{both3}(b) and (e) or
$\varepsilon=\Gamma$ in Fig.\ref{both3}(c) and (f), taking the case
of $t_1=0.1\Gamma$ as an example, the contribution of $t_1$ to the
conductance is obvious since the differences between the conductance
of $\phi=0$ and $\phi=2\pi$. Namely, in the low or high energy
regimes, the widths of the antiresonance valleys when $\phi=0$ are
different from those in the case of $\phi=2\pi$, respectively; in
addition, when $\phi=2\pi$ the space between the two antiresonant
points is smaller compared with the results in the case of $\phi=0$.
Furthermore, with the increase of $t_1$, in the absence of magnetic
flux the antiresonance point in the high-energy regime goes to
vanish and the antiresonance appears at the position of
$\varepsilon_0=-\Gamma$, whereas in the case of $\phi=2\pi$
antiresonance in the low-energy regime disappears and the
antiresonance holds at the point of $\varepsilon_0=\Gamma$.
Therefore, it can be concluded that independent of the QD levels,
when the double-level impurity couples to both QDs uniformly, the
conductance zero is associated with the antibonding state of the
impurity when $\phi=0$ but when the magnetic phase factor is tuned
to $\phi=2\pi$ the conductance zero emerges in the case of the
bonding state of impurity identical with the Fermi level.
\par

\begin{figure}
\centering \scalebox{0.35}{\includegraphics{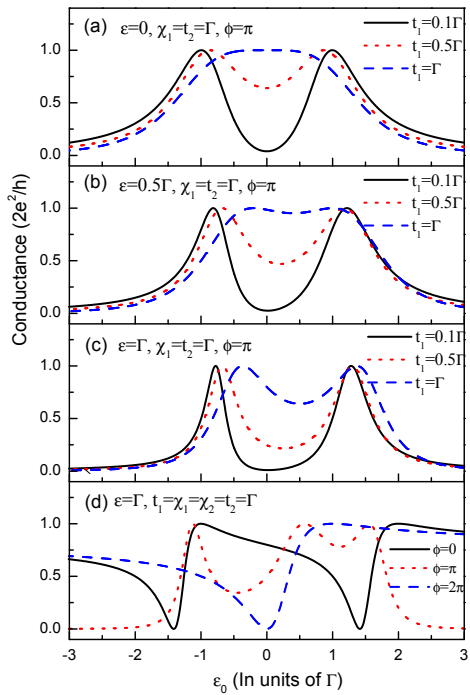}}
\caption{(a)-(c), the conductance curves with a nonlocal
double-level impurity coupled to the QDs when $\phi=\pi$. In (d),
the spectra about the triple-level nonlocal impurity are shown.
\label{both4}}
\end{figure}

Finally, with respect to the case of the magnetic flux phase factor
$\phi=\pi$, via the results in Fig.\ref{both4}(a)-(c) it is found
that despite the nonzero $t_1$ the appearance of double-level
impurity also gives rise to two resonant peaks in the conductance
spectra, while in this case no antiresonance comes about between the
two conductance peaks, but the increase of $t_1$ can efficiently
elevate the conductance valley. Combining with the above results, we
can further understand the influence of finite-value $t_1$ on the
quantum interference in this structure. When a triple-level impurity
is considered, as shown in Fig.\ref{both4}(d), the effect of
nonlocal multi-level impurity can be completely uncovered. It can be
anticipated that under the condition of the uniform couplings of
multi-level nonlocal impurity to both QDs, when $\phi=0$ the
conductance zeros are associated with the odd-numbered
(even-numbered) eigenlevels of the odd-level (even-level) impurity
but when $\phi=2\pi$ the opposite phenomenon will come into
being.\cite{jpcm}
\par
\begin{figure}
\centering \scalebox{0.35}{\includegraphics{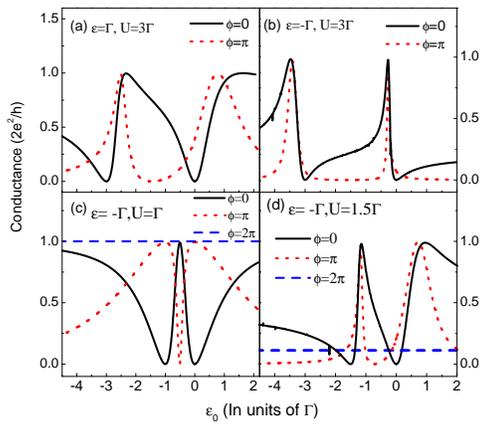}}\caption{The
nonlocal-impurity--modified linear conductance spectra in the
presence of the many-body terms with $U=3\Gamma$ in (a) and (b). In
(c) and (d), $U=\Gamma$ and $1.5\Gamma$ are considered.
\label{manybody2}}
\end{figure}

With regard to the many-body effect in such a case of nonlocal
impurity, we take the example of single-level impurity to show the
change of conductance spectra. From Fig.\ref{manybody2} with the
identical QD-impurity couplings, we can find that when Coulomb
repulsions are sufficiently strong, analogous to the results of
local impurity, the spectra have the opportunity to be divided into
two groups. And also, in the cases of $\varepsilon=-\Gamma$ the
electron interactions present remarkable effect on the electron
transport. However, in such a case the Coulomb interactions do not
lead to the disappearance or inversion of the Fano lineshapes. By
employing the results of Eq.(\ref{fanoform2}), one can understand
this phenomenon. It is of course that under the condition of the
nonlocal couplings, the Fano parameter $q_{\sigma}$ are modulated by
the impurity level as well, then, mathematically the presence of
Coulomb interactions influences the signs and values of the levels
of QDs and impurity ( $\varepsilon_1$ and $\epsilon_1$ ), which
weakens the effect of Coulomb interactions on the Fano parameter.

\section{summary}
In conclusion, in this paper we have theoretically investigated the
quantitative effect of the local and nonlocal impurities on the
electron transport through a double-QD AB interferometer, by
assuming impurities to couple to the QDs in the interferometer arms.
By plotting the linear conductance spectra \emph{vs} the impurity
levels, we have found that at the zero-magnetic-flux case, the
conductance spectra show up as the Fano lineshapes even if the
presence of a single-level impurity, but the impurity-caused
Breit-Wigner lineshapes emerge in the conductance spectra when the
magnetic flux phase factor $\phi=\pi$. Besides, the nonlocal
coupling of impurities to the QDs changes the period of conductance
change \emph{vs} the magnetic flux from $2\pi$ to $4\pi$. These
indicate the nontrivial influence of impurities on the electron
transport in this system. Furthermore, it has been seen that for the
cases of two identical impurities respectively coupled to the QDs
with uniform couplings or an impurity coupled to both QDs uniformly,
the Fano antiresonances are only relevant to the impurity levels,
including the more interesting transmission behaviors at the
nonlocal-impurity cases. By considering the many-body effect within
the second-order approximation, we also found the important role of
the Coulomb interaction in such an electron transport process. All
these results are analyzed in detail, and we expect that the
discussion in such a work can be helpful for the relevant
experiment.
\par
In the end, we would like to discuss the impurity-driven
spin-dependent electron transport when the Rashba interaction is
taken into account. Based on the theoretical description, when the
Rashba interaction is applied to the vicinity of either QD, e.g.,
QD-1, there will be a spin-dependent phase factor $\sigma\varphi$
added to the QD-lead coupling coefficients $V_{1\alpha,\sigma}$,
then, the interplay of the magnetic and Rashba fields makes the
conductance spectra of the opposite-spin electron separated from
each other\cite{Sun,Chi}. Typically, when the strengths of these two
fields are assumed to be $\phi={\pi\over2}$ and
$\varphi={\pi\over4}$, respectively, the up-spin electron thereby
undergoes a $\pi$--phase-factor quantum interference while the
down-spin electron undergoes a zero--phase-factor quantum
interference, so the apparent spin polarization occurs in the
electron transport process. Of course, the impurity can not
contribute to the occurrence of spin polarization. Here, we readily
emphasize that in the cases of different-property interferences, the
impurity plays different roles, i.e., the suppressing or enhancement
of electron transmission, so the impurity can efficiently invert the
spin polarization when the Rashba interaction is introduced, and
such an effect will be more clear when multi-level impurity is
considered.

\clearpage

\bigskip

\end{document}